\documentclass[prd,twocolumn,showpacs,amsmath,amssymb]{revtex4-1}
\usepackage[utf8]{inputenc}
\usepackage{amsmath}
\usepackage{latexsym}
\usepackage{amsfonts}
\usepackage{graphicx}
\usepackage{mathrsfs}
\usepackage{CJK}
\usepackage{longtable}

\usepackage{hyperref}
\hypersetup{
  colorlinks = true,
  urlcolor = blue,
  linkcolor = blue,
  citecolor = green,
  filecolor = magenta,
}

\newcommand{\ud}{\mathrm{d}}

\begin{document}
\begin{CJK*}{GB}{song}

\title{Gravitational Lensing of Gravitational Waves: Rotation of Polarization Plane}
\author{Shaoqi Hou}
\affiliation{School of Physics and Technology, Wuhan University, Wuhan, Hubei 430072, China}
\author{Xi-Long Fan}
\email{xilong.fan@whu.edu.cn}
\affiliation{School of Physics and Technology, Wuhan University, Wuhan, Hubei 430072, China}
\author{Zong-Hong Zhu}
\email{zhuzh@whu.edu.cn}
\affiliation{School of Physics and Technology, Wuhan University, Wuhan, Hubei 430072, China}
\date{\today}

\begin{abstract}
Similar to the light, gravitational waves traveling in multiple paths may arrive at the same location if there is a gravitational lens on their way.
Apart from the magnification of the amplitudes and the time delay between the gravitational wave rays, gravitational lensing also rotates their polarization planes and causes the gravitational wave Faraday rotation.
The effect of the Faraday rotation is weak and can be ignored.
The rotation of the polarization plane results in the changes in the antenna pattern function, which describes the response of the detector to its relative orientation to the gravitational wave.
These effects are all reflected in the strain, the signal registered by the interferometers.
The gravitational wave rays in various directions stimulate different strains, mainly due to different magnification factors, the phases and the rotation of the polarization plane.
The phase difference mainly comes from the time delay.
Moreover, the rotation of the polarization plane seemingly introduces the \textit{apparent} vector polarizations, when these strains are compared with each other.
Because of the smallness of the deflection angles, the effect of the rotation is also negligible.
\end{abstract}

\maketitle
\end{CJK*}

\section{Introduction}

The detection of the 11 gravitational wave (GW) events by LIGO/Virgo collaborations \cite{Abbott:2016blz,Abbott:2016nmj,Abbott:2017vtc,Abbott:2017oio,TheLIGOScientific:2017qsa,Abbott:2017gyy,LIGOScientific:2018mvr} confirmed the prediction of Einstein's general relativity (GR) \cite{Einstein:1916cc,Einstein:1918btx}, and marked the new era of GW astronomy and astrophysics. 
Among them, GW170817, together with GRB 170817A, verified that GWs in GR are traveling at the speed of light \cite{TheLIGOScientific:2017qsa,Goldstein:2017mmi,Savchenko:2017ffs,Monitor:2017mdv}.
So GWs would experience the similar gravitational lensing by the gravitational potential on their way \cite{Lawrence1971nc,Lawrence:1971hx,Ohanian:1974ys,Takahashi:2003ix}.
In the geometrical optics  regime, their  trajectories would bend, and then might come together at the Earth, enabling the detection of all of them.   
Finally, the amplitudes of these GWs change by different factors due to the focusing effect of the lensing.

In the gravitational lensing of light, strong lensing is referred to the case where the images of the lensed object can be distinguished by the observer.
Einstein rings or arcs are also within the regime of strong lensing. 
If the distortions of the images are much smaller, the gravitational lensing is said to be weak \cite{Paolis:2016roy}.
Similar concepts might be defined for gravitational lensing of GWs, where ``images'' cannot be seen, but heard. 
Strong lensing of GWs can be used to put constraints on modified theories of gravity, formation and evolution of structures, and the Hubble constant $H_0$ \cite{Sereno:2010dr,Sereno:2011ty}.
The cosmology and the speed of the GW can also be constrained with the strong lensing of both GW and light \cite{Liao:2017ioi,Fan:2016swi}.
The primordial dark matter power spectrum and the growth of structure can be constrained using the weak lensing of GWs based on the method in Ref.~\cite{Cutler:2009qv}.
The extension of the previous method could discriminate different cosmological models ($\Lambda$CDM, dynamical dark energy/quintessence, modified gravity \textit{et al.}) \cite{Camera:2013xfa}.
These models can also be constrained with the future joint inference of standard siren and GW weak lensing \cite{Congedo:2018wfn}.
In this work, we primarily consider the strong GW lensing.

There are also some differences between the GW and the light.
For example, GWs have much longer wavelengths than the light.
GWs, generated by a binary star system and traveling in various directions, are monochromatic with varying frequencies in time and differ by definite phases from each other \cite{Maggiore:1900zz,Poisson2014}, so they are coherent waves. 
Usually, they arrive at the detector at well-separated times, and are measured individually; if the earth nearly lines up with the lens and the binary stars, it is possible to observe the lensed GWs simultaneously, which might produce interesting phenomena of beat, the interference between GWs \cite{beatprl}. 
Other wave effects of GW could be found in \cite{Nakamura:1997sw,Nakamura1999wo,Takahashi:2003ix,Liao:2019aqq}.  
In contrast, lights, emanating from a star, possess random frequencies and phases, so are incoherent.    

And the third difference is related to the methods of detection, or the variables used to describe the light and the GW.
For the light, it is the intensity that is usually measured, which is a scalar quantity, so does not fully explore the vector nature of the light.
To the contrary, interferometers measure the strain, the relative change in the arm lengths.
The strain, or rather the antenna pattern function (APF) \cite{Isi:2015cva}, which is basically the strain for GWs with the unit amplitude, depends on the polarization of the GW.
Therefore, it reflects the tensorial nature of GWs, which is the main target of this work.

GWs in GR have two polarizations, the well-known plus (+) and cross ($\times$) polarizations \cite{Misner:1974qy}.
These polarizations are transverse, meaning the GW oscillates in the plane perpendicular to the direction of propagation.
This plane can be called the polarization plane and is parallel transported along the null geodesic of the GW. 
When the null trajectory bends due to the presence of a gravitational lens, the polarization plane rotates accordingly.
It is thus expected that the APF for the lensed GW will be different from the one for the GW that is not lensed. 
Likewise, the APFs for the lensed GWs in different paths will also differ from each other, depending on their directions.

The rotation of the polarization plane was not considered in the previous studies on the gravitational lensing of GWs \cite{Lawrence1971nc,Lawrence:1971hx,gravlens1992,ArnaudVarvella:2003va,Dai:2016igl,Fan:2016swi,Ng:2017yiu,Dai:2018enj}, except a short remark in a footnote in Ref.~\cite{Ohanian:1974ys}.
In the studies of the wave nature of GWs in gravitational lensing \cite{Takahashi:2003ix,Liao:2019aqq}, the tensor aspects of GWs is completely ignored in order to solve the Kirchhoff's equation easily. 
Effectively, they treated the GW as a scalar field. 
In this work, this rotation effect will be specifically studied, discussing how the APF changes and what contributes to the modified strain in the gravitational lensing caused by a Newtonian potential.
Since the Newtonian potential is very weak, the deflection angle is small, so this effect can be safely ignored in most situations, 
which will be clarified in the following discussion.

The difference in APFs or the strains for different GWs could be explained by the gravitational lensing.
However, it might also be simply explained by the possibility that GWs come from sources at distinct locations but from almost the same direction.
So there  seems to be a degeneracy. 
One way to resolve the degeneracy is to realize that gravitational lensing barely affects the frequency evolution of the signal. 
In fact, the frequency evolution of the observed GW \textit{does} change due to the inhomogeneous matter distribution along the path of the GW as well as the expansion of the universe according to Refs.~\cite{Seto:2001qf,Nishizawa:2011eq,Bonvin:2016qxr}.
However, interferometers have difficulties in detecting this effect, as the frequency shift is generally small for ground-based detectors and slightly larger for LISA \cite{Bonvin:2016qxr,Seoane:2013qna}. 
Moreover, the mismatch would be at most $10^{-3}$ if a template without this effect considered were used for LISA. 
In addition, the recent paper \cite{Tamanini:2019usx} concludes that LISA cannot measure the peculiar acceleration of a binary black hole system, which causes the aforementioned change in the frequency evolution, during the nominal 4 years' run.
Therefore, from the point of view of the observation, gravitational lensing basically does not affect how the frequency of the detected GW varies over the course of detection.
Based on this, the statistics method provided in Ref.~\cite{Haris:2018vmn} can be used to tell whether some GW events detected by the LIGO/Virgo network are lensed or not. 
So in this work, we will also concentrate on the lensed GW events that could be detected by the LIGO/Virgo network.

This work is organized in the following way. 
Section~\ref{sec-pro} reviews how the GW propagates in a generic spacetime background in the geometric optics limit; three propagation effects will be discussed, including the rotation of the polarization plane and the GW Faraday rotation.
Section~\ref{sec-glgw} discusses how the polarization plane rotates due to the gravitational lensing caused by a point mass. 
After that, the strain of the lensed GW is calculated in Sec.~\ref{sec-mea}. 
Finally, a brief discussion and  conclusion are given in Sec.~\ref{sec-dc}.
Appendix~\ref{sec-app} derives a general formula to calculate the strain.
Throughout this work, the geometrized units ($G=c=1$) are used.
Round brackets enclosing indices imply symmetrization, while square brackets mean antisymmetrization, e.g., $T_{(\mu\nu)}=(T_{\mu\nu}+T_{\nu\mu})/2$, and $T_{[\mu\nu]}=(T_{\mu\nu}-T_{\nu\mu})/2$.

\section{The propagation of gravitational waves} 
\label{sec-pro}

In the short wavelength limit, the Einstein's equation determines three major properties of GWs.
First, GWs propagate in null geodesics.
Second, the polarization tensors are parallel transported along the trajectories, and the number of gravitons is conserved; since the direction of the propagation changes, the polarization plane also rotates \cite{Ohanian:1974ys}. 
Third, the GW Faraday rotation occurs at high enough orders \cite{Piran1985nf,Piran:1985dk,Wang:1991nf}.
In the following, these three properties will be analyzed in order. 

GWs are perturbations $h_{\mu\nu}$ to the spacetime metric $g_{\mu\nu}=g^B_{\mu\nu}+h_{\mu\nu}$ with $g^B_{\mu\nu}$ describing the background geometry.
When the wavelength $\lambda$ of the GW is much smaller than the characteristic curvature radius $\mathcal R$ of the background geometry,  the geometric optics limit applies.
In the transverse-traceless (TT) gauge ($\nabla^\nu\bar h_{\mu\nu}=0$ and $g_\text{B}^{\mu\nu}\bar h_{\mu\nu}=0$), it satisfies the following perturbed vacuum Einstein's equation \cite{Isaacson:1967zz},
\begin{equation}\label{eq-ein-p}
  \nabla_\rho\nabla^\rho\bar h_{\mu\nu}+2R_{\mu\rho\nu\sigma}^B\bar h^{\rho\sigma}=0,
\end{equation}
where $\bar h_{\mu\nu}=h_{\mu\nu}-g^B_{\mu\nu}g_B^{\rho\sigma}h_{\rho\sigma}/2$ is the trace-reversed perturbation, $\nabla_\mu$ is the covariant derivative compatible with $g^B_{\mu\nu}$ and $R^B_{\mu\rho\nu\sigma}$ is its curvature tensor.
In the following, the raising and the lowering indices are done with $g^\text{B}_{\mu\nu}$ and its inverse $g_\text{B}^{\mu\nu}$.
One can write the GW as $\bar h_{\mu\nu}=\Re[(A_{\mu\nu}+\epsilon B_{\mu\nu}+\cdots)e^{i\Phi/\epsilon}]$ with $\Re$ standing for the real part \cite{Misner:1974qy}.
Here, $\epsilon$ is a formal expansion parameter which indicates that the terms multiplied by $\epsilon^n$ are of the order of $(\lambda/\mathcal R)^n$.
The traceless condition leads to $A^\mu{}_\mu=B^\mu{}_\mu=0$.
Define $l_\mu=-\nabla_\mu\Phi$.
At the leading orders $O(1/\epsilon^2)$ and $O(1/\epsilon)$, Eq.~\eqref{eq-ein-p} gives  $l_\mu l^\mu=0$, and
\begin{gather}
  l^\rho\nabla_\rho A_{\mu\nu}+\frac{1}{2}A_{\mu\nu}\nabla_\rho l^\rho=0,\label{eq-sj-2}
\end{gather}
and the Lorenz gauge is equivalent to $A_{\mu\nu}l^\nu=0$.
Since $\nabla_\mu l_\nu=-\nabla_\mu\nabla_\nu \Phi=\nabla_\nu l_\mu$, one also finds out that $l^\nu\nabla_\nu l^\mu=0$.
Indeed, in the geometric optics limit, $l^\mu$ is null and parallel transported along its own integral curve.

GWs have polarizations \cite{Will:2014kxa,Gong:2018ybk}.
To express the polarizations in a convenient way, one associates the graviton with a tetrad basis $\{l^\mu,\,n^\mu,\, x^\mu,\, y^\mu\}$, which are parallel transported along the trajectory of the graviton, and satisfy $-l^\mu n_\mu= x^\mu x_\mu= y^\mu y_\mu=1$ with the remaining contractions vanishing.
For convenience, the Newman-Penrose (NP) null tetrad $\{l^\mu,n^\mu,m^\mu,\bar m^\mu\}$ can also be used with $m^\mu=(x^\mu-iy^\mu)/\sqrt{2}$ and $\bar m^\mu=(x^\mu+iy^\mu)/\sqrt{2}$, which is suitable for describing radiations \cite{Newman:1961qr,Stephani:2003tm}. 
 
For the two GW polarizations in GR, their polarization tensors are defined as
\begin{equation}\label{eq-def-pols}
e^+_{\mu\nu}= x_\mu x_\nu- y_\mu y_\nu,\quad
e^\times_{\mu\nu}= x_\mu y_\nu+ y_\mu x_\nu.
\end{equation}
These tensors are also parallel transported along the trajectory of the graviton, i.e., 
\begin{equation}
  \label{eq-pte}
  l^\rho\nabla_\rho e^P_{\mu\nu}=0,\quad P=+,\times.
\end{equation}
Usually, one chooses a gauge such that $\bar h_{0\mu}=0$, and the spatial components $e^P_{ij}$ are often used to represent the polarization tensors \cite{Will:2014kxa,Isi:2015cva}.
Equation~\eqref{eq-pte}, together with Eq.~\eqref{eq-sj-2}, leads to the evolution of the amplitudes,
\begin{equation}\label{eq-ev-am}
  l^\mu\nabla_\mu A^P+\frac{1}{2}A^P\nabla_\mu l^\mu=0.
\end{equation}
This shows that the leading order polarizations evolve separately and in exactly the same way.
Although the amplitudes vary along the trajectories of gravitons,  the numbers of gravitons are constant, i.e., $\nabla_\mu[|A^P|^2l^\mu]=0$.

The gauge invariant quantities describing the GW are some components of the Weyl tensor $C_{\mu\nu\rho\sigma}$, i.e. the NP variable $\Psi_4=C_{\mu\nu\rho\sigma}n^\mu\bar m^\mu n^\rho\bar m^\sigma$ \cite{Newman:1961qr,Stephani:2003tm}.
A short calculation gives
  \begin{equation}\label{eq-psi4-1}
    \Psi_4^{(1)}=\frac{1}{2}\Re(A^+e^{i\Phi})+\frac{i}{2}\Re(A^\times e^{i\Phi}),
  \end{equation}
at the order $O(1/\epsilon)$. 
There is a freedom to choose the NP tetrad.
In particular, one can carry out a spin, i.e., $l'^\mu=l^\mu$, $n'^\mu=n^\mu$ and $m'^\mu=e^{i\varphi}m^\mu$ for some angle $\varphi$, then $\Psi_4$ transforms according to  \cite{Chandrasekhar:1985kt}
\begin{equation}
  \Psi_4\rightarrow \Psi'_4=e^{-i2\varphi}\Psi_4.
\end{equation}
The angle $\varphi=\varphi_+$ at which $\Im(\Psi'_4)=0$ is called the polarization angle \cite{Wang:1991nf},  given by
\[
  \varphi_+=\frac{1}{2}\arctan\frac{\Im(\Psi_4)}{\Re(\Psi_4)},
\] 
with $\Im$ representing the imaginary part.
At the order [$O(1/\epsilon)$],  
\[
  \varphi_+^{(1)}=\frac{1}{2}\arctan\frac{\Re (A^\times)}{\Re (A^+)}.
\]
Due to Eq.~\eqref{eq-ev-am}, $l^\mu\nabla_\mu\varphi_+^{(1)}=0$, so there is no GW Faraday rotation  at this order $O(1/\epsilon)$.  
The GW Faraday rotation has been known for a while, but here, we examine this effect in the context of the gravitational lensing for the first time.

At the next  order $O(1)$ , one finds out that \cite{Misner:1974qy}
\begin{equation}
l^\rho\nabla_\rho B_{\mu\nu}+\frac{1}{2}B_{\mu\nu}\nabla_\rho l^\rho=\frac{i}{2}\nabla_\rho\nabla^\rho A_{\mu\nu}+iR^B_{\mu\rho\nu\sigma}A^{\rho\sigma}, \label{eq-sb-2}
\end{equation}
which  shows that the evolution of $B_{\mu\nu}$ is affected by $A_{\mu\nu}$ as well as the background geometry.
This feature also appears in even higher order corrections to $A_{\mu\nu}$, which is discussed in Ref.~\cite{Harte:2018wni}.
This behavior would generally lead to the GW Faraday rotation \cite{Piran1985nf,Piran:1985dk,Wang:1991nf}.
The Lorenz gauge condition for $B_{\mu\nu}$ reads $B_{\mu\nu}l^\nu=i\nabla^\nu A_{\mu\nu}$, so $B_{\mu\nu}$ is not transverse to $l^\mu$.
In the following, the focus will be on the propagation of GWs in the background generated by a Newtonian potential (the lens).
As the Newtonian potential is very weak, and $B_{\mu\nu}$ is at least of second order in the Newtonian potential, so it will be ignored completely.

\section{Gravitational lensing of gravitational waves} 
\label{sec-glgw}

We only work in the geometrical optical  regime in this work.
Consider the gravitational lensing caused by a Newtonian potential $M/r$.
After passing the lens, the deflected 4-velocity of the GW is approximately $l^\mu=(1,\vec{\boldsymbol{l}}+\vec\alpha)$ in the limit where the lens is far away from both the source of the GW and the detector.
Here, $\vec{\boldsymbol{l}}$ is the original direction of the GW, and $\vec\alpha=-4M\vec b/b^2$ is the deflection vector \cite{Poisson2014}, where $\vec b$ is the impact vector, which is perpendicular to $\vec{\boldsymbol{l}}$ and whose magnitude is the distance of the closest approach.
One also determines the remaining of the tetrad basis, which are 
\begin{gather}
  n^\mu=\frac{1}{2}(1,-\vec{\boldsymbol{l}}),\\
x^\mu=\left(\frac{1}{2}\vec\alpha\cdot\vec{\boldsymbol{x}},\vec{\boldsymbol{x}}-\frac{1}{2}(\vec\alpha\cdot\vec{\boldsymbol{x}})\vec{\boldsymbol{l}}\right),\\
y^\mu=\left(\frac{1}{2}\vec\alpha\cdot\vec{\boldsymbol{y}},\vec{\boldsymbol{y}}-\frac{1}{2}(\vec\alpha\cdot\vec{\boldsymbol{y}})\vec{\boldsymbol{l}}\right).
\end{gather}
Here, a set of orthonormal 3-vectors $\{\vec{\boldsymbol{l}}, \vec{\boldsymbol{x}}, \vec{\boldsymbol{y}}\}$ (triads) is introduced, satisfying $\vec{\boldsymbol{l}}=\vec{\boldsymbol{x}}\wedge\vec{\boldsymbol{y}}$.

When the GW is emitted and far away from the lens, assume that the components of the trace-reversed metric perturbation $\bar h_{\mu\nu}$ are $\bar h_{0\mu}=0$ and $\bar h_{ij}=\bar A_{ij}e^{i\Phi}$, where $\bar A_{ij}=\bar A^+\bar e^+_{ij}+\bar A^\times\bar e^\times_{ij}$ with
\begin{equation}
 \bar e^+_{ij}={\boldsymbol{x}}_i{\boldsymbol{x}}_j-{\boldsymbol{y}}_i{\boldsymbol{y}}_j,\quad \bar e^\times_{ij}={\boldsymbol{x}}_i{\boldsymbol{y}}_j+{\boldsymbol{y}}_i{\boldsymbol{x}}_j. 
\end{equation} 
$\bar e^+_{ij}$ and $\bar e^\times_{ij}$ would be the polarization matrices if there were no lens.
For a binary system of two stars with masses $m_1$ and $m_2$, circling around each other in an orbit of radius $a$, the amplitudes during the inspiral phase are approximately given by \cite{Poisson2014},
\begin{gather}
\bar A^+=\mathcal{A}\left[-\frac{1+\cos^2\iota}{2}\cos2\psi+i\cos\iota\sin2\psi\right],\label{eq-ap-o}\\
\bar A^\times=\mathcal{A}\left[i\cos\iota\cos2\psi+\frac{1+\cos^2\iota}{2}\sin2\psi\right],\label{eq-ac-o}
\end{gather}
where  $\mathcal A=\frac{4m_1 m_2}{aR}e^{-i2\varpi}$, $R$ is the distance from the source to the observer, $\psi$ is the polarization angle and $(\iota,\,\varpi)$ represents the angular direction of $\vec{\boldsymbol{l}}$ in the source frame with $\iota$ actually the inclination angle.
From these expressions, one finds out that the amplitudes also depend on the direction ($\iota,\varpi$) and the polarization angle $\psi$ of the GW.
The contribution of the angle $\varpi$ can be absorbed into the phase $\Phi$.
So the GWs emanating from the binary star system in various directions not only have different amplitudes, but also differ in the initial phase.

\begin{figure}
  \centering
  \includegraphics[width=0.45\textwidth]{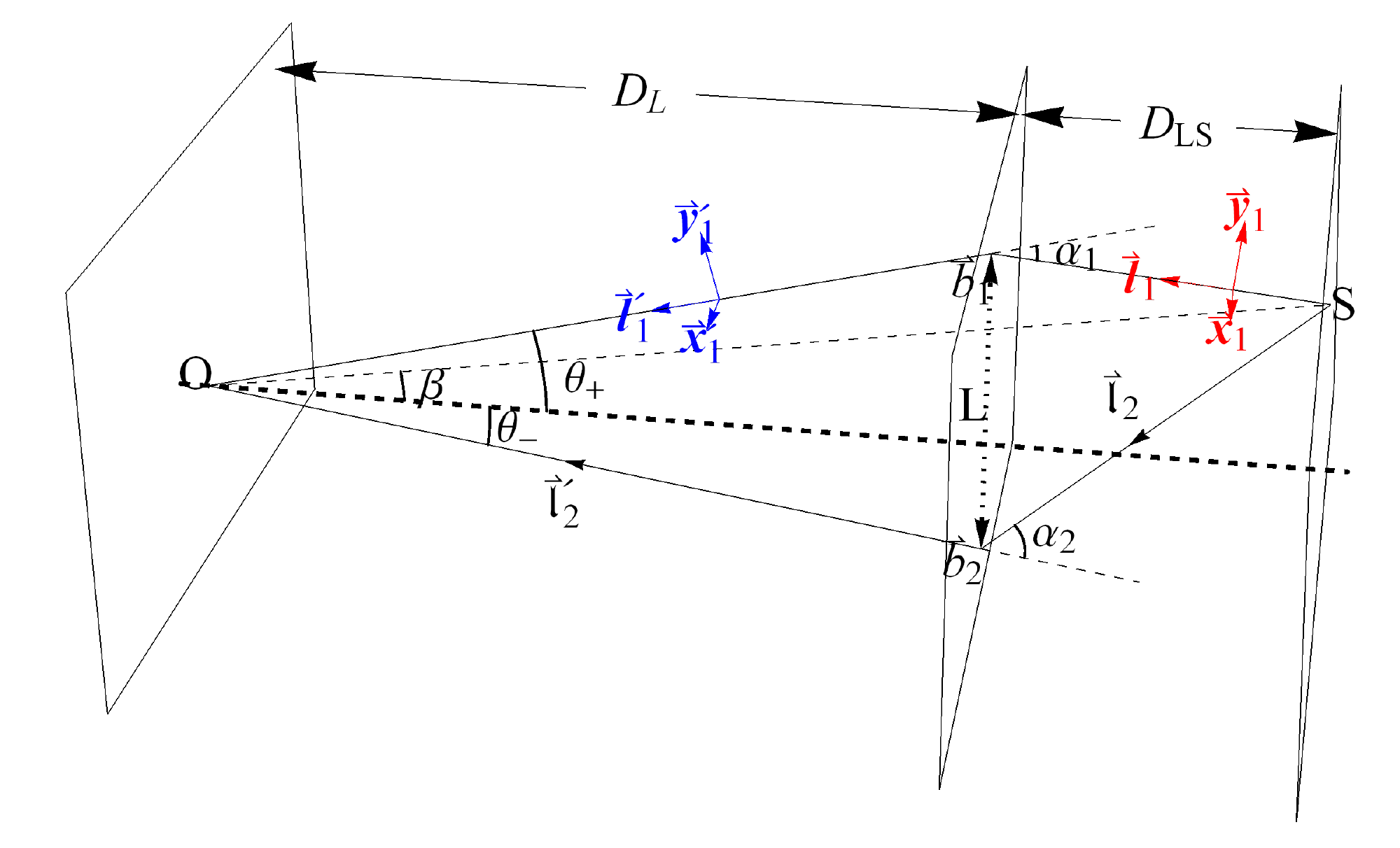}
  \caption{Geometry of a Schwarzschild lens.
  S is the position of the binary star system, and the interferometer is at O.
  L represents the gravitational lens, and the thick dashed line is the optical axis.
  The vertical squares represent the observer, lens and source planes, from the left to the right.
  $\beta$ is the misalignment angle between the optical axis and the line connecting O to S.
  Two GW rays 1 and 2 are emitted from S, initially in the directions of $\vec{\boldsymbol{l}}_1$ and $\vec{\boldsymbol{l}}_2$, respectively.
  After passing the lens plane, their directions change, given by $\vec{\boldsymbol{l}}'_1$ and $\vec{\boldsymbol{l}}'_2$, forming angles $\theta_\pm$ with the optical axis. 
  The deflection angles are $\alpha_1$ and $\alpha_2$, respectively.
  }\label{fig-geogl}
\end{figure}

For the Schwarzschild lens considered, there will be two ``images" of the source, which are located at the angles \cite{gravlens1992}  
\begin{equation}
  \label{eq-thepm}
  \theta_\pm=\frac{\beta\pm\sqrt{\beta^2+4\theta_\text{E}^2}}{2},
\end{equation} 
as shown in Fig.~\ref{fig-geogl}.
Here, $\theta_\text{E}=\sqrt{\frac{4M}{c^2}\frac{D_\text{LS}}{D_\text{S}D_\text{L}}}$ is the Einstein angle, and if the cosmological evolution is considered, one has  \cite{gravlens1992}
\begin{gather}
  (1+z_\text{L})\frac{D_\text{L}D_\text{S}}{D_\text{LS}}=\frac{1}{\chi(z_\text{L})-\chi(z_\text{S})}, \\
  \chi(z)=\int_z^{+\infty}\frac{\ud\zeta}{H(\zeta)D^2(\zeta)(1+\zeta)^2},
\end{gather} 
where, in the denominator of the integrand, $H(z)$ is the Hubble parameter at the redshift $z$ and $D(z)$ is the angular diameter distance 
, so $D_\text{S}=D(z_\text{S})$ and $D_\text{L}=D(z_\text{L})$ are the angular diameter distances of the source and the lens, respectively.
In this work, we mainly focus on the case where $z\lesssim 2$ because according to Ref.~\cite{Yang:2019jhw} the probability for lensed GWs from neutron star-neutron star mergers peaks around $z=2$, while the probability for lensed GWs from black hole-black hole mergers peaks around $z=4$.
Because of the focusing effect of the lens, the amplitudes of  are enhanced by
\begin{equation}
  \label{eq-mf}
{\mu_\pm}=\frac{|\theta_\pm|}{\sqrt{|\theta_+^2-\theta_-^2|}}, 
\end{equation}
respectively.
The time delay between the two rays is 
\begin{equation}
  \Delta t  =4M(1+z_\text{L})\left(\frac{\theta_+^2-\theta_-^2}{2\theta_\text{E}^2}+\ln\frac{\theta_+}{-\theta_-}\right),
\end{equation}
which contributes partially to the phase difference,
\begin{equation}\label{eq-pd-td}
  \Delta\Phi=\omega\Delta t.
\end{equation}

Now, after passing by the gravitational lens, the polarization tensors for the GW  are  given by 
\begin{equation}
  e^P_{\mu\nu}=\left(
  \begin{array}{cc}
    0 & -\displaystyle\frac{1}{2}\bar e^P_{ik}\alpha^k \\
    -\displaystyle\frac{1}{2}\bar e^P_{jk}\alpha^k & \bar e^P_{ij}+\tilde e^P_{ij}
  \end{array}\right),\quad P=+,\times,\label{eq-ep}
\end{equation}
according to Eq.~\eqref{eq-def-pols}, where $\tilde e^P_{ij}$  are the corrections to $\bar e^P_{ij}$ given by
\begin{gather}
\tilde e^+_{ij}=-\frac{1}{2}(\vec\alpha\cdot\vec{\boldsymbol{x}})\bar e^{x}_{ij}+\frac{1}{2}(\vec\alpha\cdot\vec{\boldsymbol{y}})\bar e^{y}_{ij},\\
\tilde e^\times_{ij}=-\frac{1}{2}(\vec\alpha\cdot\vec{\boldsymbol{y}})\bar e^{x}_{ij}-\frac{1}{2}(\vec\alpha\cdot\vec{\boldsymbol{x}})\bar e^{y}_{ij}.
\end{gather}
Finally, in the above expressions, $\bar e^{x}_{ij}=\boldsymbol{l}_i{\boldsymbol{x}}_j+{\boldsymbol{x}}_i \boldsymbol{l}_j$ and $\bar e^{y}_{ij}=\boldsymbol{l}_i{\boldsymbol{y}}_j+{\boldsymbol{y}}_i \boldsymbol{l}_j$ are the vector polarization matrices  for the unperturbed GW.
Therefore, after passing a gravitational lens, the GW changes its direction of motion.
Since the polarization tensors $e^P_{\mu\nu}$ are parallel transported, they are also modified \cite{Ohanian:1974ys}.

Note that  the  gravitational lensing inducing vector polarizations ($\bar e^{x}_{ij}$ and $\bar e^{y}_{ij}$) is an illusion.
In fact, the \textit{appearance}  of $\bar e^{x}_{ij}$ and $\bar e^{y}_{ij}$ is simply due to the use of the original triads $\{\vec{\boldsymbol{l}},\,\vec{\boldsymbol{x}},\,\vec{\boldsymbol{y}}\}$ to describe the changed GW polarizations, not because of the existence of some vector degrees of freedom as in certain alternative metric theories of gravity \cite{Gong:2018cgj,Gong:2018vbo,Gong:2018ybk,Hou:2018djz}.
The gravitational lensing causes the rotation of the propagation vector of the GW, from $\vec{\boldsymbol{l}}$  to $\vec{\boldsymbol{l}}'=\vec{\boldsymbol{l}}+\vec\alpha$.
Note that neither $\vec{\boldsymbol{x}}$ nor $\vec{\boldsymbol{y}}$ is perpendicular to $\vec{\boldsymbol{l}}'$.
This can be remedied by adding to $ x^\mu$ and $ y^\mu$ some linear combinations of $l^\mu$, $ x^\mu$ and $ y^\mu$, for example, $x'^\mu= x^\mu-(\vec\alpha\cdot\vec{\boldsymbol{x}}) l^\mu/2=(0,\vec{\boldsymbol{x}}-(\vec\alpha\cdot\vec{\boldsymbol{x}})\vec{\boldsymbol{l}})$ and $y'^\mu= y^\mu-(\vec\alpha\cdot\vec{\boldsymbol{y}}) l^\mu/2=(0,\vec{\boldsymbol{y}}-(\vec\alpha\cdot\vec{\boldsymbol{y}})\vec{\boldsymbol{l}})$.
Now, call
\begin{gather}
\vec{\boldsymbol{x}}'=\vec{\boldsymbol{x}}-(\vec\alpha\cdot\vec{\boldsymbol{x}})\vec{\boldsymbol{l}},\quad
\vec{\boldsymbol{y}}'=\vec{\boldsymbol{y}}-(\vec\alpha\cdot{\boldsymbol{y}})\vec{\boldsymbol{l}}.\label{eq-ntps-xy1}
\end{gather}
The use of the (primed) triads $\{\vec{\boldsymbol{l}}',\,\vec{\boldsymbol{x}}',\,\vec{\boldsymbol{y}}'\}$ to represent the GW polarizations will not introduce the apparent vector polarizations.
To sum up, the appearance of the vector polarizations is simply because one expresses the polarization matrices in terms of the original (unprimed) triads.
There are still two tensor polarizations.

Equation \eqref{eq-ntps-xy1}  shows that the triad gets rotated by a small angle, after the GW passes the lens.
This effect is displayed in Fig.~\ref{fig-geogl}, where the red triad represents the initial basis $\{\vec{\boldsymbol{l}},\vec{\boldsymbol{x}},\vec{\boldsymbol{y}}\}$, and the blue one represents the final basis $\{\vec{\boldsymbol{l}}',\vec{\boldsymbol{x}}',\vec{\boldsymbol{y}}'\}$ for the GW ray 1.
In order to calculate the strain, one has to compute the APFs first, which is the topic of the next section.

\section{The measurement  of  lensed gravitational waves} 
\label{sec-mea}

When the GW reaches the interferometers, it causes the change in the lengths of the arms.
This kind of the response of the detector is quantified by the so-called antenna pattern function \cite{Isi:2015cva}.
To calculate this function, one needs compute the Riemann tensor of the GW.

According to Ref.~\cite{Harte:2018wni}, the leading order of the Riemann tensor for the GW is given by 
\begin{equation}
 R^\text{GW}_{\mu\nu\rho\sigma}=-2\omega^2e^{i\Phi}l_{[\mu}A_{\nu][\rho}l_{\sigma]}, 
\end{equation}
with $\omega$ the frequency of the GW.
So the electric part of it is 
\begin{equation}
 R^\text{GW}_{tjtk}=\sum_{P=+,\times}\frac{\omega^2}{2}e^{i\Phi}(2 e^P_{0(j}\boldsymbol l_{k)}+ e^P_{jk}). 
\end{equation}
Let $D^{jk}=\frac{1}{2}(\hat X^j\hat X^k-\hat Y^j\hat Y^k)$ represent the configuration of an interferometer, with unit vectors $\hat X^j$ and $\hat Y^j$ pointing in the directions of the arms.
The strain is thus given by 
\begin{equation}
  \label{eq-st-int}
h(t)=-2D^{jk}\int\ud t\int\ud t'R^\text{GW}_{tjtk},
\end{equation}
whose justification is relegated in the Appendix~\ref{sec-app}.
Performing the double integration and dropping the factor of the amplitude and the phase, the antenna pattern functions are simply given by
\begin{gather}
  F^+=\bar F^+-\frac{1}{2}(\vec\alpha\cdot\vec{\boldsymbol{x}})\bar F^{x}+\frac{1}{2}(\vec\alpha\cdot\vec{\boldsymbol{y}})\bar F^{y},\label{eq-atp-+}\\
  F^\times=\bar F^\times-\frac{1}{2}(\vec\alpha\cdot\boldsymbol{y})\bar F^{x}-\frac{1}{2}(\vec\alpha\cdot\vec{\boldsymbol{x}})\bar F^{y},\label{eq-atp-x}
\end{gather}
for the ground-based detectors, where $\bar F^P=D^{ij}\bar e^P_{ij}$ are the antenna pattern functions for the unperturbed GW.
$F^{P}$ depend not only on the unperturbed polarization matrices $\bar e^P_{ij}$, but also the corrections $\tilde e^{P}_{ij}$.
These relations show that the APFs get modified by the gravitational lensing.

Now, compare the APFs for the two GW rays 1 and 2.
The initial triads for these GW rays are $\{\vec{\boldsymbol{l}}_1,\,\vec{\boldsymbol{x}}_1,\,\vec{\boldsymbol{y}}_1\}$ and $\{\vec{\boldsymbol{l}}_2,\,\vec{\boldsymbol{x}}_2,\,\vec{\boldsymbol{y}}_2\}$ \footnote{Not shown in Fig.~\ref{fig-geogl}, otherwise it would be too cumbersome.}, respectively.
Let $\vec\delta=\vec{\boldsymbol{l}}_1-\vec{\boldsymbol{l}}_2$, whose magnitude is $\delta=\arccos(\vec{\boldsymbol{l}}_1\cdot\vec{\boldsymbol{l}}_2)$, a small angle.
Then, $\vec{\boldsymbol{x}}_2$ and $\vec{\boldsymbol{y}}_2$ can be approximately expressed as $\vec{\boldsymbol{x}}_2=\vec{\boldsymbol{x}}_1+(\vec\delta\cdot\vec{\boldsymbol{x}}_1)\vec{\boldsymbol{l}}_1$ and $\vec{\boldsymbol{y}}_2=\vec{\boldsymbol{y}}_1+(\vec\delta\cdot\vec{\boldsymbol{y}}_1)\vec{\boldsymbol{l}}_1$ up to an arbitrary rotation around $\vec{\boldsymbol l}_1$.
One can thus relate the polarization tensors of the second GW to those of the first one in the following way,
\begin{eqnarray}\label{eq-pols-21}
  \bar e^+_{2,ij}&=&\bar e^+_{1,ij}+(\vec\delta\cdot\vec{\boldsymbol x}_1)\bar e^{x}_{1,ij}-(\vec\delta\cdot\vec{\boldsymbol{y}}_1)\bar e^{y}_{1,ij},\\
  \bar e^\times_{2,ij}&=&\bar e^\times_{1,ij}+(\vec\delta\cdot\vec{\boldsymbol x}_1)\bar e^{y}_{1,ij}+(\vec\delta\cdot\vec{\boldsymbol{y}}_1)\bar e^{x}_{1,ij}.
\end{eqnarray}
Therefore, the relations between the APFs for the two gravitons are,
\begin{eqnarray}\label{eq-apts-21}
  F^+_2&=& F_1^++(\vec\delta'\cdot\vec{\boldsymbol{x}}_1) F_1^{x}-(\vec\delta'\cdot\vec{\boldsymbol{y}}_1) F_1^{y},\\
  F^\times_2&=&F^\times_1+(\vec\delta'\cdot\vec{\boldsymbol{x}}_1) F_1^{y}+(\vec\delta'\cdot\vec{\boldsymbol{y}}_1) F_1^{x},
\end{eqnarray}
where $F_1^{x}\approx D^{ij}\bar e^{x}_{1,ij}$, $F_1^{y}\approx D^{ij}\bar e^{y}_{1,ij}$, and $\vec\delta'\equiv\vec{\boldsymbol{l}}'_1-\vec{\boldsymbol{l}}'_2=\vec\delta-(\vec\alpha_2-\vec\alpha_1)$.

In an ideal situation, one can measure the angle $\delta'=-(\theta_+-\theta_-)$, where the overall minus indicates that $\vec\delta'$ points downward in Fig.~\ref{fig-geogl}.
Since the deflections $\alpha_1$ and $\alpha_2$ are small, it is a good approximation that the vectors $\vec\delta'$, $\vec\delta$, $\vec\alpha_1$, and $\vec\alpha_2$ are parallel to each other.
Therefore, $\delta'=\delta+\alpha_1-\alpha_2$.
From Fig.~\ref{fig-geogl}, one recognizes that $b_1=\theta_+D_\text{L}$, $b_2=\theta_-D_\text{L}$, and $b_1b_2/D_\text{L}^2=\theta_+\theta_-=-\theta_\text{E}^2$, so $\delta=\left(1-\frac{4M}{\theta_\text{E}^2D_\text{L}}\right)\delta'$.
In this way, the initial and the final angular separations of the two GW rays are related.
As long as $\delta$ is known, it is possible to infer the initial phase difference between the rays 1 and 2.

To fully appreciate the effects of the gravitational lensing, the strains for the two GW rays 1 and 2 should be compared.
The strain caused by the GW 2 is related to that of the GW 1 in the following way,
\begin{equation}\label{eq-s2-s1}
\begin{split}
  h_2=&A_2^+F_2^++A_2^\times F_2^\times\\
  \approx&e^{-i\Delta\Phi}\frac{\mu_-}{\mu_+}\bigg\{h_1+\left[(\vec\delta\cdot\vec{\boldsymbol{x}}_1)A_1^{x}-(\vec\delta\cdot\vec{\boldsymbol{y}}_1)A_1^{y}\right]F_1^+\\
  &+\left[(\vec\delta\cdot\vec{\boldsymbol{x}}_1)A_1^{y}+(\vec\delta\cdot\vec{\boldsymbol{y}}_1)A_1^{x}\right]F_1^\times\\
  &+A_1^+\left[(\vec\delta'\cdot\vec{\boldsymbol{x}}_1)F_1^{x}-(\vec\delta'\cdot\vec{\boldsymbol{y}}_1)F_1^{y}\right]\\
  &+A_1^\times\left[(\vec\delta'\cdot\vec{\boldsymbol{x}}_1)F_1^{y}+(\vec\delta'\cdot\vec{\boldsymbol{y}}_1)F_1^{x}\right]\bigg\},
  \end{split}
\end{equation}
where $h_1=A_1^+F_1^++A_1^\times F_1^\times$ with $A_1^{P}\approx{\mu_{+}}\bar A_1^P$ [refer to Eqs.~\eqref{eq-ap-o} and \eqref{eq-ac-o}], and the phase difference $\Delta\Phi$ is given by Eq.~\eqref{eq-pd-td}.
$A_1^{x}$ and $A_1^{y}$ are approximately
\begin{gather}
\bar A^{x}=\mathcal A\left[-\frac{1}{2}\sin2\iota\cos\psi+i\sin\iota\sin\psi\right],\\
\bar A^{y}=\mathcal A\left[\frac{1}{2}\sin2\iota\sin\psi+i\sin\iota\cos\psi\right],
\end{gather}
multiplied by ${\mu_+}$ and evaluated along the GW ray 1 at the observer, respectively.

The differences in the strains $h_1$ and $h_2$ are multiple.
First, they differ from each other in phase, which comes from (1) the time delay $\Delta t$ [refer to Eq.~\eqref{eq-pd-td}], and (2) propagation direction in the source frame, i.e., $\varpi$ in Eqs.~\eqref{eq-ap-o} and \eqref{eq-ac-o}.
Second, the magnification factors  (${\mu_\pm}$) are not the same.
Third, the polarization planes underwent distinct rotations, which is the reason for the existence of the terms in the curly brackets except $h_1$.

\section{Discussion and conclusion} 
\label{sec-dc}

For a lens of mass $(10^6 - 10^{12})M_\odot$ and $\beta\sim1$ arcsecond, the deflection angle $\alpha=2M/b$ is about 1 arcsecond.
Further decreasing the misalignment angle $\beta$ causes even smaller $\alpha$.   
Although  we only work in the geometrical optical  regime in this work, for the most cases,  the terms with $\delta$ and $\delta'$ in Eq.~\eqref{eq-s2-s1} are smaller than $h_1$ by at least 6 orders of magnitude, so they can be safely ignored. 
This justifies the ignorance of the rotation of the polarization plane in Refs.~\cite{Takahashi:2003ix,Liao:2019aqq}, although these works discussed the wave nature of the GW.
In addition, Refs.~\cite{Lawrence1971nc,Lawrence:1971hx,gravlens1992,ArnaudVarvella:2003va,Dai:2016igl,Fan:2016swi,Ng:2017yiu,Dai:2018enj} neglected the effects of the rotation because the authors mainly considered lensed signals that are well separated in time. 
The particularly interesting situation where there is a time window when the lensed signals are simultaneously observed is discussed in Ref.~\cite{beatprl}.

The difference between $h_1$ and $h_2$ mainly comes from the magnification and the phase shift caused by the time delay. 
The ratio $\mu_-/\mu_+$ between the magnification factors of the two GW rays can be much smaller than 1, especially when the misalignment angle $\beta$ is large, which can be estimated as 
\begin{equation}
  \frac{\mu_-}{\mu_+}=-\frac{\theta_-}{\theta_+}\approx\frac{(\theta_\text{E}/\beta)^2}{1+(\theta_\text{E}/\beta)^2},
\end{equation} 
by Eqs.~\eqref{eq-thepm} and  \eqref{eq-mf}.
The magnification factors are very close to each other when $\beta\ll1$ arcsecond.
In this case, the geometric optics might not be sufficient to describe the lensing.
The phase difference $\Delta\Phi$ is very huge as the time delay could ranges from a few days to a few months.
So in the geometric optics regime, the change in the strain is dominated by the magnification and the phase shift.

Finally, the GW Faraday rotation is also one interesting phenomena, which is due to the interaction between the GW and the background geometry \cite{Piran1985nf,Piran:1985dk,Wang:1991nf}.
However, it happens at higher orders in the short-wavelength limit, so unlikely be observed in the near future. \\

\begin{acknowledgements}
This work was supported by the National Natural Science Foundation of China under Grants Nos. 11633001 and the Strategic Priority Research Program of the Chinese Academy of Sciences, Grant No. XDB23000000.
\end{acknowledgements}

\appendix

\section{Strain}
\label{sec-app}

In this Appendix, Eq.~\eqref{eq-st-int} is derived.
One starts with the linearized geodesic deviation equation \cite{Misner:1974qy}
\begin{equation}\label{eq-gdv}
  \ddot x^j=-R_{tjtk}^\text{GW}x^k,
\end{equation}
where $x^j$ represents the relative displacement between two test particles, e.g., the mirrors used in the interferometers.
Integrating twice gives the change in the relative displacement,
\begin{equation}\label{eq-cinx}
  \delta x^j=-x_0^k\int\ud t\int\ud t'R_{tjtk} ^\text{GW},
\end{equation}
where $x_0^k$ stands for the initial relative displacement.
Here, one assumes that the total relative displacement $x^j(t)$ remains the same, i.e., $x^j(t)=x_0^j$, which is a good approximation as the GW is weak.

Now, consider the effect of the GW on an interferometer whose arms have the initial length $L$, and are in the directions $\hat x^j_1$ and $\hat x^j_2$.
The change in the length of the first interferometer arm in the direction given by $\vec x_1/L=\hat x_1$ is
\begin{equation}\label{eq-cinl}
  \delta L_1=\delta\sqrt{\delta_{jk}x_{1}^jx^k_1}=-\frac{x_1^jx_1^k}{L}\int\ud t\int \ud t'R_{tjtk}^\text{GW}.
\end{equation}
One can obtain the similar expression for the second arm.
Then the strain is
\begin{equation}\label{eq-stn-g}
  h(t)=\frac{\delta L_1-\delta L_2}{L}=-2D^{jk}\int\ud t\int \ud t'R_{tjtk}^\text{GW},
\end{equation}
where $D^{jk}=(\hat x_1^j\hat x_1^k-\hat x_2^j\hat x^k_2)/2$.
From the derivation, one understands that the above expressions applies to any metric theory of gravity in any gauge.
Since the angle $\arccos(\hat x_1\cdot\hat x_2)$ between the two arms is not specified, this result also applies to Einstein Telescope \cite{Punturo:2010zza}.
In the transverse-traceless (TT) gauge, $R_{tjtk}^\text{GW}=-\ddot h_{jk}^\text{TT}/2$, then one recovers the usual expression for the strain,
\begin{equation}
  h(t)=D^{jk}h_{jk}^\text{TT}.
\end{equation}

Equation~\eqref{eq-st-int} can thus be obtained by setting $\hat x_1=\hat X$ and $\hat x_2=\hat Y$.
This justifies the correctness of Eq.~\eqref{eq-st-int}.

\bibliographystyle{apsrev4-1}
\bibliography{rotatepol_v3_fan.bbl}

\end{document}